\documentclass[twocolumn,showpacs,preprintnumbers,amsmath,amssymb]{revtex4}

\usepackage{graphicx}% Include figure files
\usepackage{dcolumn}% Align table columns on decimal point
\usepackage{bm}% bold math
\usepackage{subfigure}

\begin{document}

\preprint{}

\title{Resolving Small-scale Structures in  Two-dimensional Boussinesq Convection
by Spectral Methods with High Resolutions}

\author{Z. Yin}
 \email{zhaohua.yin@imech.ac.cn}
\affiliation{National Microgravity Laboratory, Institute of
Mechanics, Chinese Academy of Sciences, Beijing 100080, PR China}

\author{Tao Tang}
 \email{ttang@math.hkbu.edu.hk}
 \affiliation{Department of Mathematics, Hong Kong Baptist
University, Kowloon Tong, Hong Kong}

\date{\today}% It is always \today, today,

\begin{abstract}
Two-dimensional Boussinesq convection is studied numerically with
very fine spatial resolutions up to
$4096^2$. Our numerical study
starts with a smooth asymmetric initial condition, which is
chosen to make the flow field well confined in the
computational domain until the blow-up time ($T_c$). Our study
shows that the vorticity will blow up at a finite time with
$|\omega|_{max} \sim {(T_c-t)}^{-1.61}$ and $|\nabla \theta|_{max}
\sim {(T_c - t)}^{-3.58}$.
\end{abstract}


\maketitle

Understanding whether smooth initial conditions in
three-dimensional (3D) Euler equations can develop singularities
in a finite time is an important step in understanding
high-Reynolds-number hydrodynamics~\cite{mar02,fri03}.
Two-dimensional (2D) Boussinesq convection is a simplified model
of the 3D axisymmetric flows with swirl if main flow structures
are well away from the symmetry axis ~\cite{pum92a}. The computing
requirements of the 2D Boussinesq simulations are significantly less
than those of the 3D Euler equations. However, the spatial
resolutions adopted have still been not fine enough to resolve the
small scale structure. Therefore, the spectral methods ~\cite{e94}
and the adaptive mesh methods ~\cite{pum92a,pum92b,cen01} are
mostly adopted. One central issue of this study is on whether
there exist singularities in the 2D Boussinesq flows. If we denote
$T_c$ the time of blowup, the minimum criterion for the breakdown
of smooth solutions in the 2D Boussinesq equations is: $
|\omega|_{max} \sim {(T_c - t)}^{-\alpha}$ and $|\nabla
\theta|_{max} \sim {(T_c - t)}^{-\beta}$, where $\alpha \geq 1 $
and $\beta \geq 2$ ~\cite{e94, cha97}. Adaptive mesh techniques
~\cite{pum92a,pum92b,cen01} and the spectral method ~\cite{e94}
are adopted to investigate this problem, which yield various
conclusions, e.g.,  ~\cite{e94,cen01} observe no vorticity
blow-up, while ~\cite{pum92a,pum92b} only provide the marginal values
($\alpha = 1$ and $\beta = 2$) although these studies predict
vorticity singularities.

Since the fast developed adaptive mesh methods are limited by the
finite-order accuracy and mesh equality, a further ``spectral''
effort seems necessary to investigate this challenging problem.
The spectral methods have been used intensively in 3D Euler
simulations, but they are limited by the available computing
capability: the finest resolutions used so far have been $2048^3$
~\cite{cic05}. The conclusions are affected by certain kind of
symmetric assumptions introduced to increase the effective
resolutions~\cite{tay35,kid85}. We can argue that the conclusion
drawn by the 3D studies~\cite{cic05,bor94} is not necessarily more
convincing than the axisymmetric assumption ~\cite{e94} where a
resolution of $1500^2$ is adopted. In this study,
spectral computations with extremely
high resolution (from $2048^2$ to $4096^2$)
are carried out to investigate the blow-up issue for the 2D
Boussinesq convection problem. Moreover, we try to maintain the flow
structure well away from the axis at the time when solutions
begin to blow up. This is done by choosing appropriate initial
data. The governing equations are
solved by a fully de-alias pseudospectral Fourier methods with 8/9
phase-shifted scheme. The digital filter is adopted to modify the
Fourier coefficient to increase the stability of the numerical
codes. The machine accuracy of our computer with double precision
is $\epsilon = 10^{-16} \approx e^{-37}$, and the modifying factor
in the filter is $\varphi(k) = e^{-37(k/N)^{16}}$ for $k < N$
~\cite{van91}. With these efforts, the vorticity blow-up is
observed, which is in contrast to the conclusion drawn by
the earlier spectral computations ~\cite{e94}.

\begin{figure*}
\begin{minipage}[c]{.28 \linewidth}
\scalebox{1}[1]{\includegraphics[width=\linewidth]{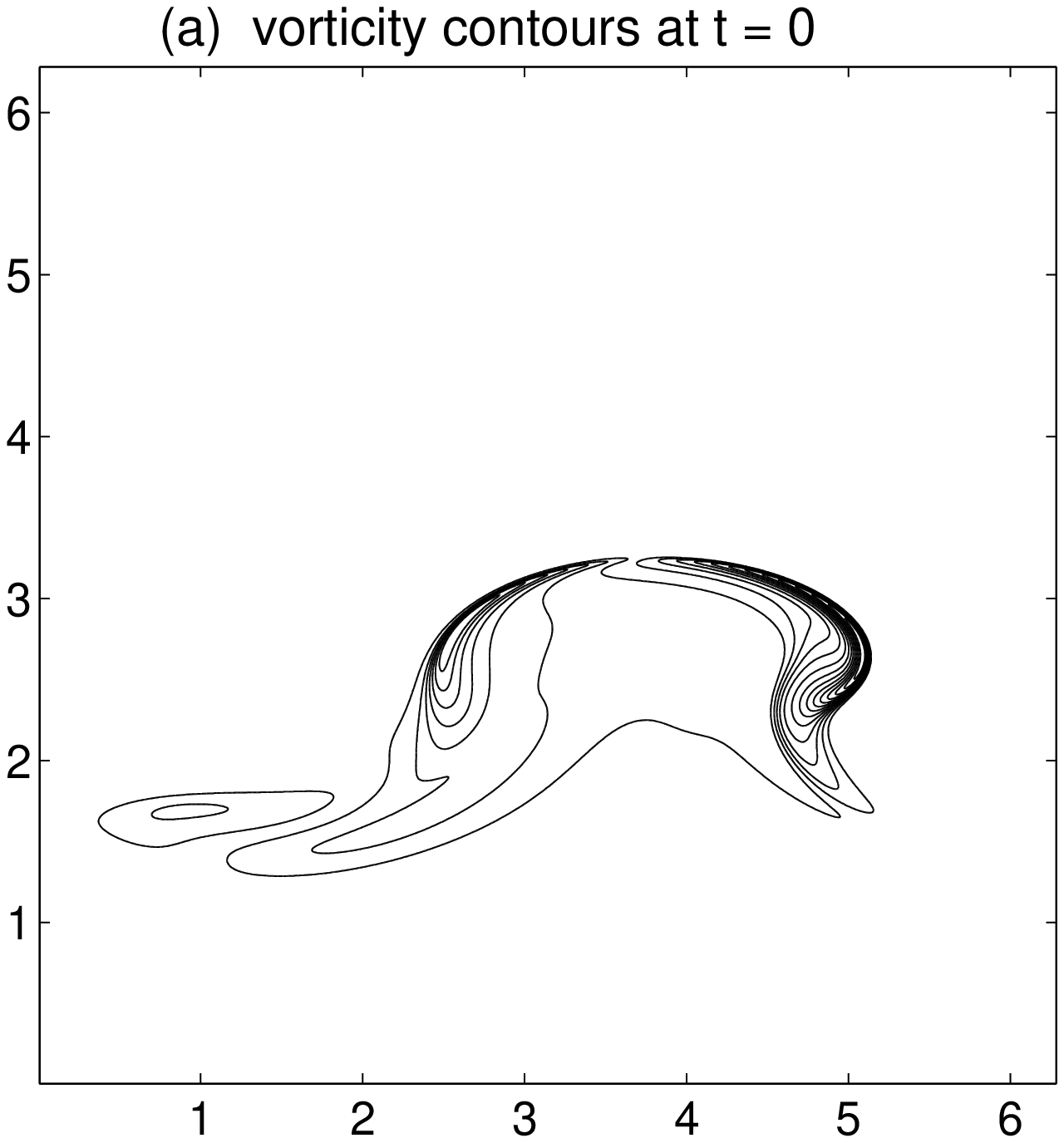}}
\end{minipage}
\begin{minipage}[c]{.28 \linewidth}
\scalebox{1}[1]{\includegraphics[width=\linewidth]{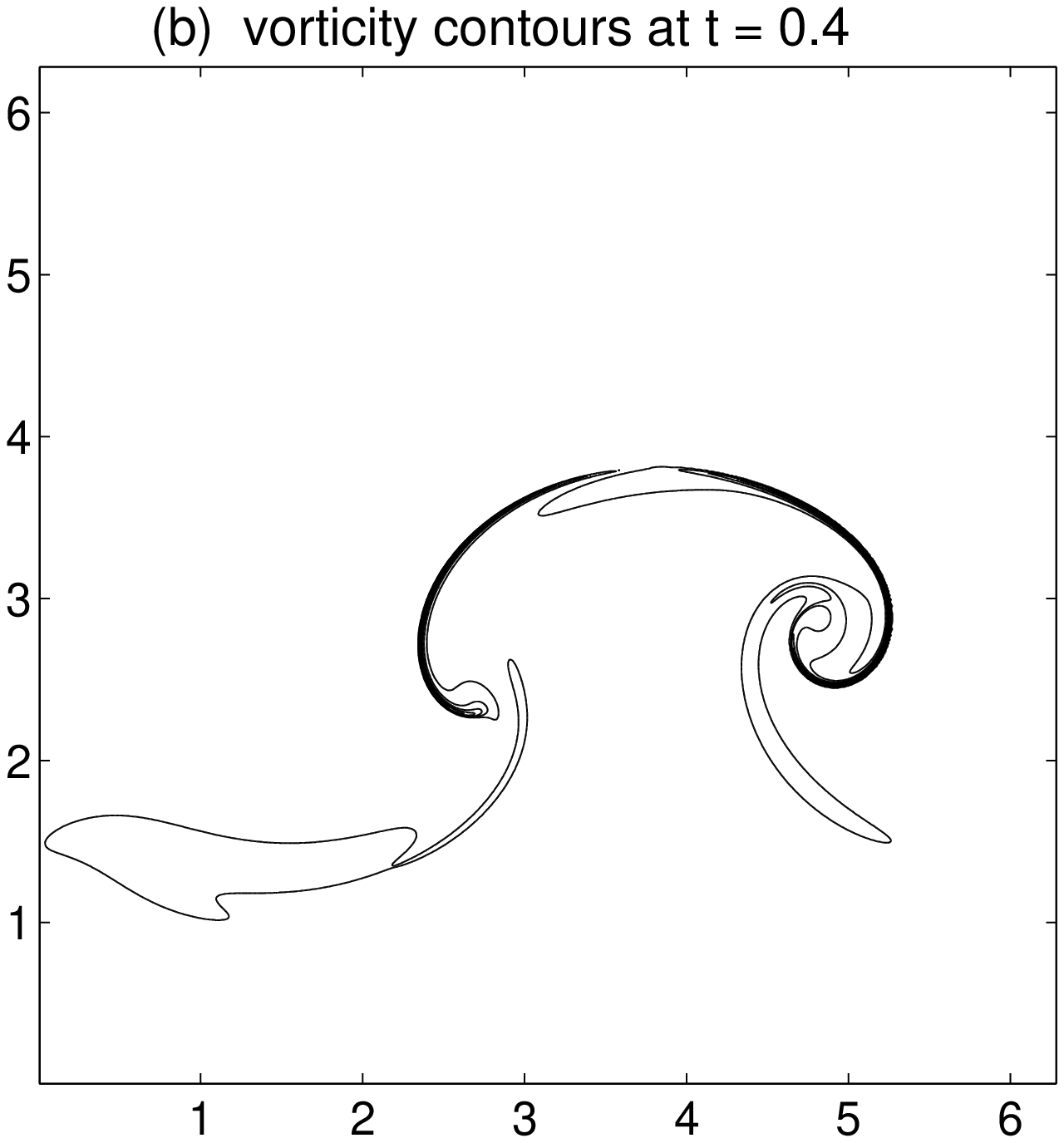}}
\end{minipage}
\begin{minipage}[c]{.28 \linewidth}
\scalebox{1}[1]{\includegraphics[width=\linewidth]{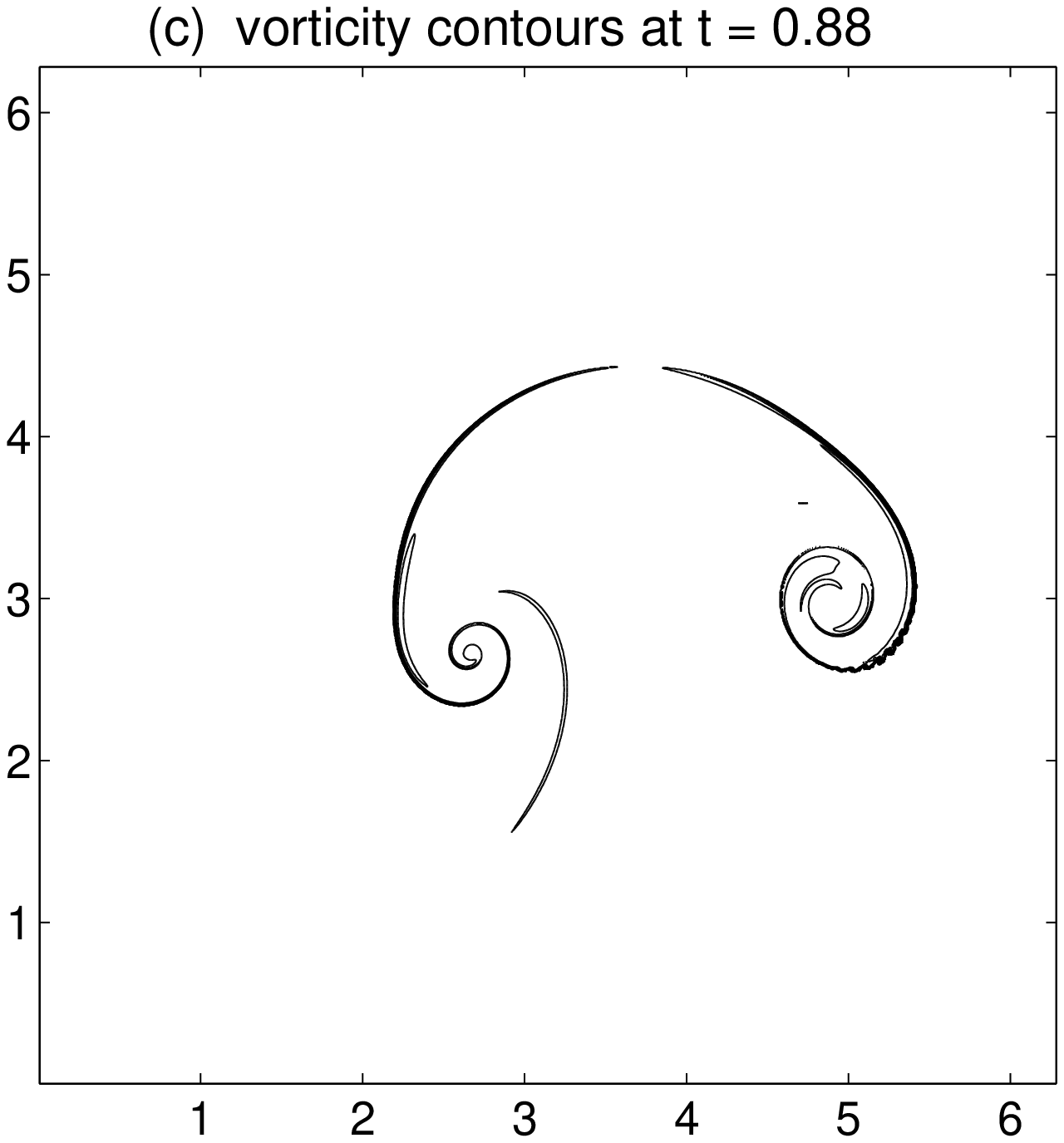}}
\end{minipage}
\begin{minipage}[c]{.28 \linewidth}
\scalebox{1}[1]{\includegraphics[width=\linewidth]{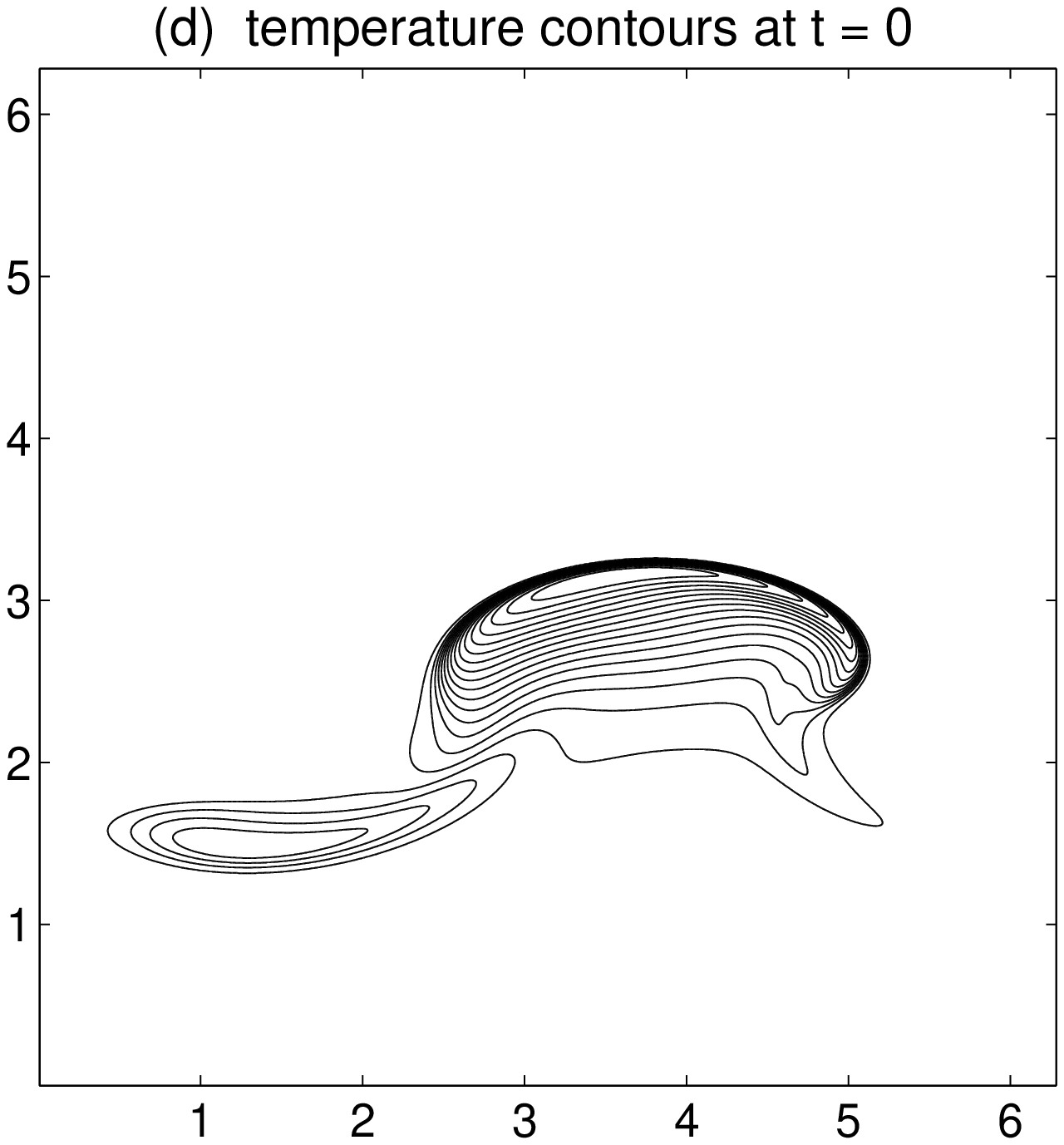}}
\end{minipage}
\begin{minipage}[c]{.28 \linewidth}
\scalebox{1}[1]{\includegraphics[width=\linewidth]{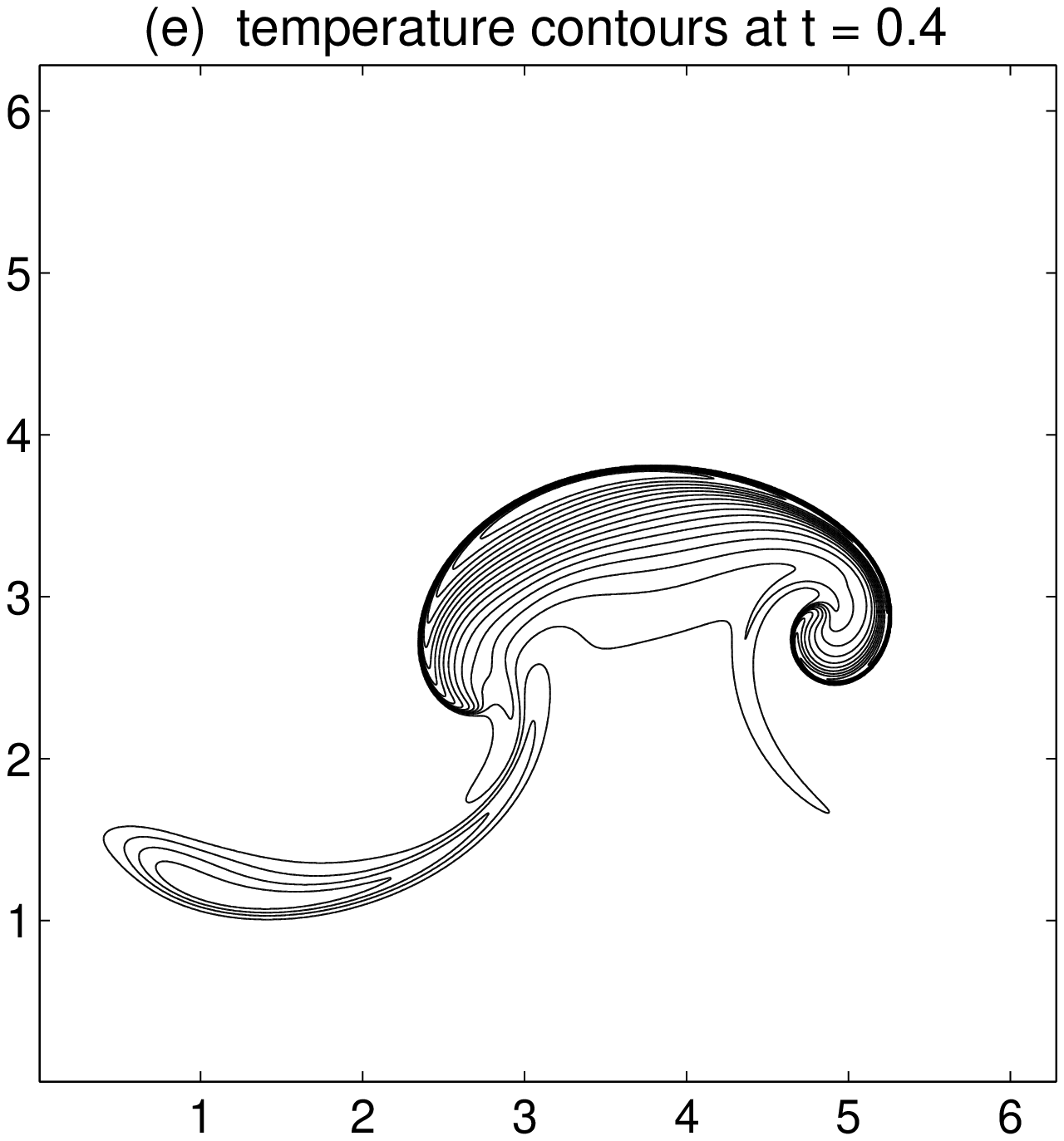}}
\end{minipage}
\begin{minipage}[c]{.28 \linewidth}
\scalebox{1}[1]{\includegraphics[width=\linewidth]{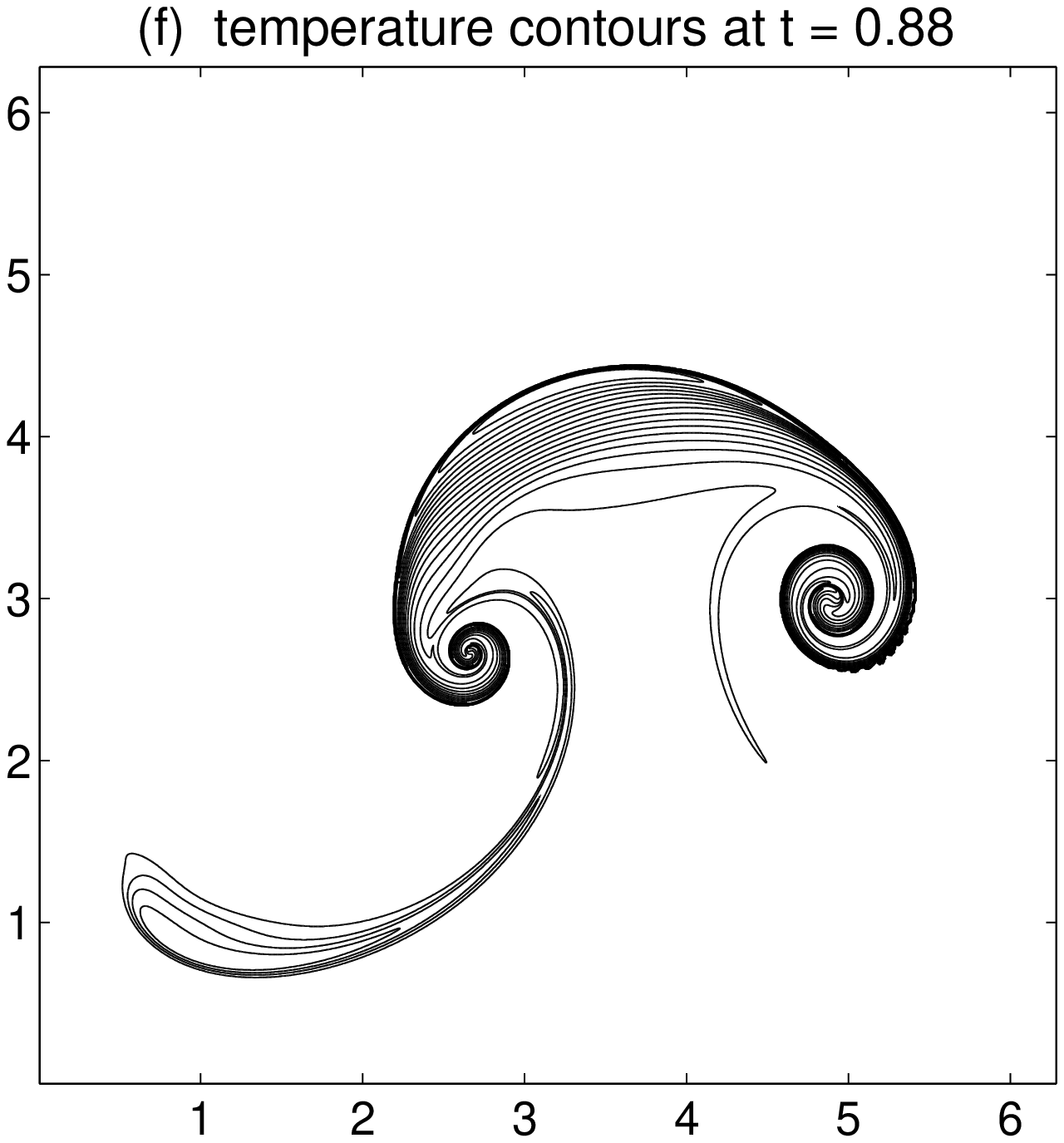}}
\end{minipage}
\caption{\label{fig:contour}Contour plots of temperature and
vorticity at different times with the resolution of $4096^2$.}
\end{figure*}

\begin{figure*}
\begin{minipage}[c]{.48 \linewidth}
\scalebox{1}[1.05]{\includegraphics[width=\linewidth]{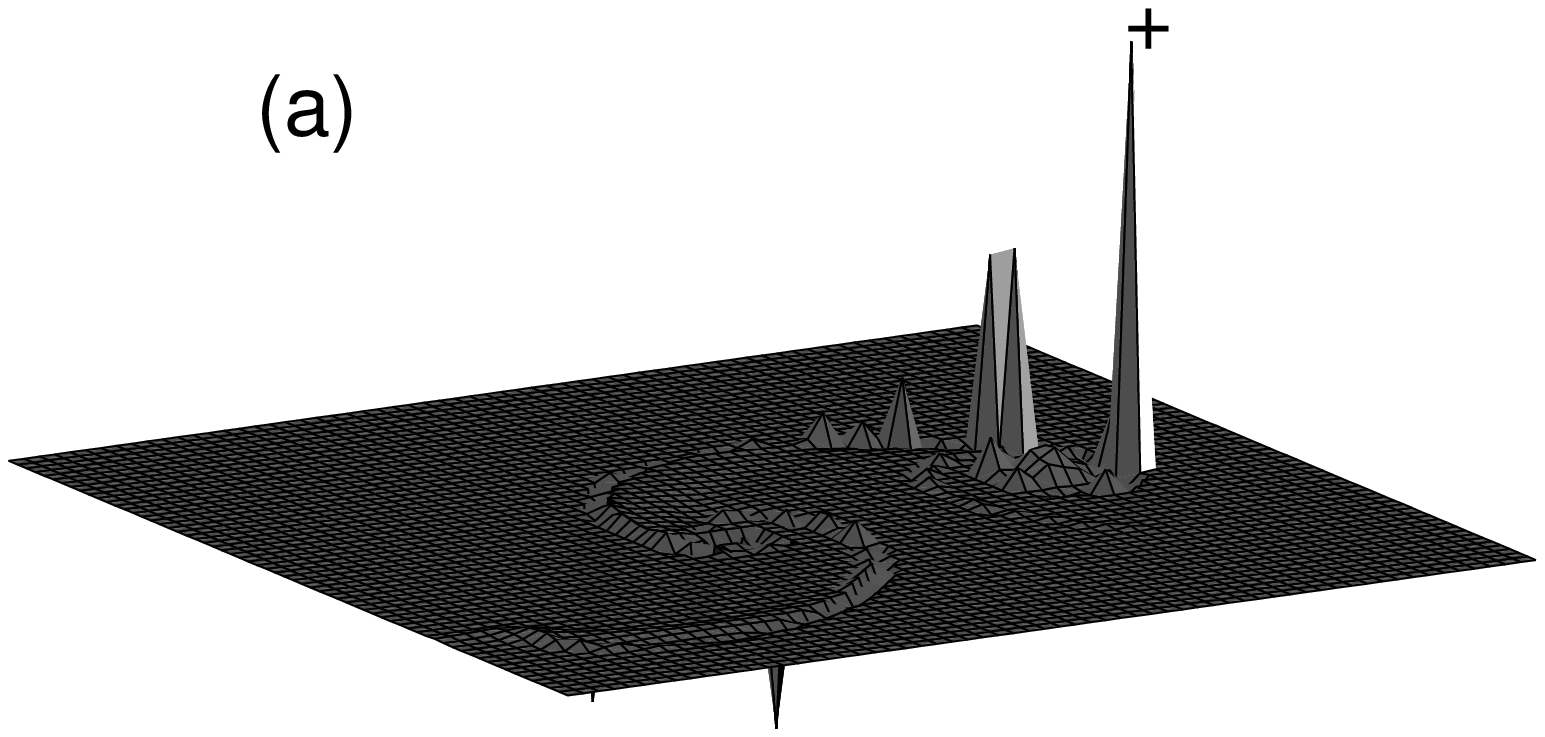}}
\end{minipage}
\begin{minipage}[c]{.44 \linewidth}
\scalebox{1}[1.]{\includegraphics[width=\linewidth]{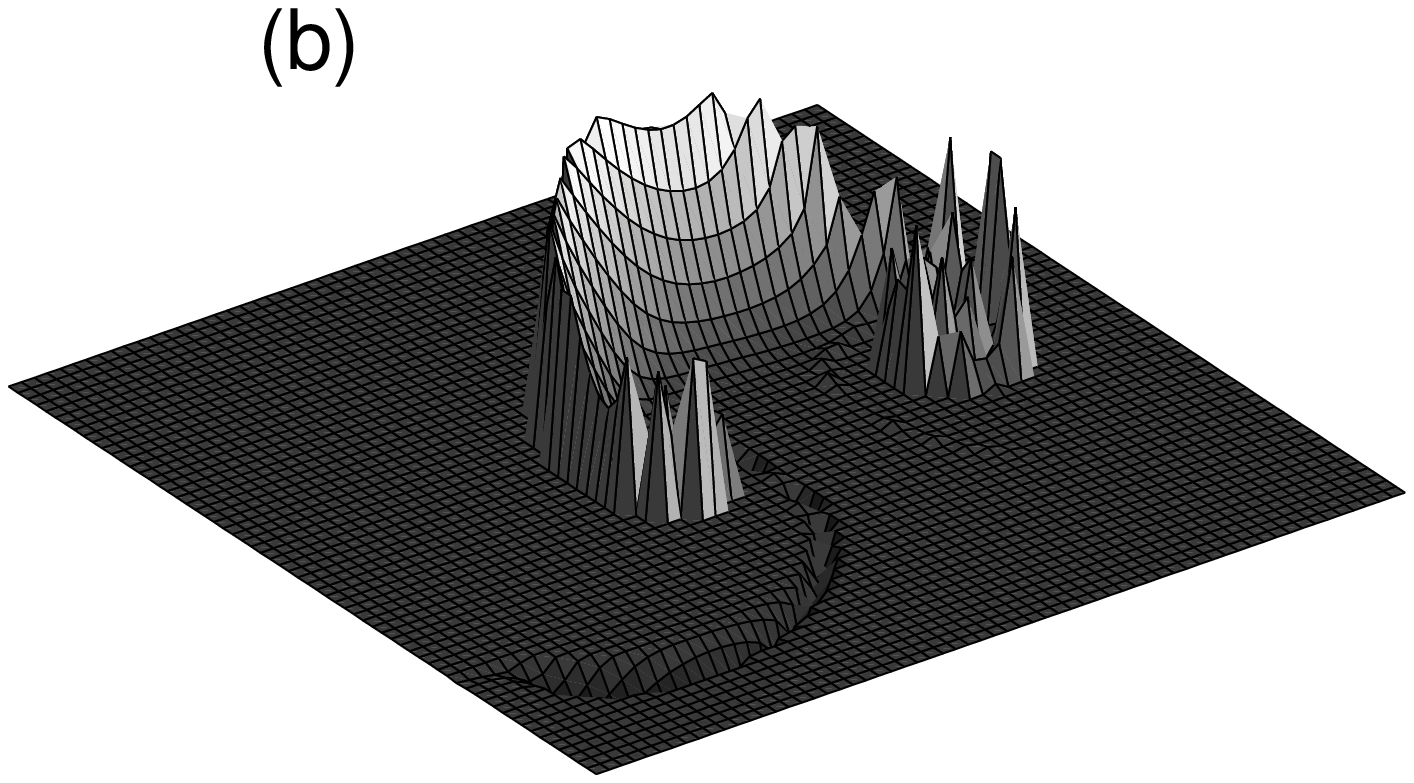}}
\end{minipage}
\caption{\label{fig:vorticity} Three-dimensional perspective plot
of the computed vorticity and temperature versus $x$ and $y$ at $t
= 0.88$ in $4096^2$ run. The vorticity peak near the ``+'' on (a)
is the location of the maximum $|\omega|$.}
\end{figure*}

\begin{figure*}
\begin{minipage}[c]{.3 \linewidth}
\scalebox{1}[1]{\includegraphics[width=\linewidth]{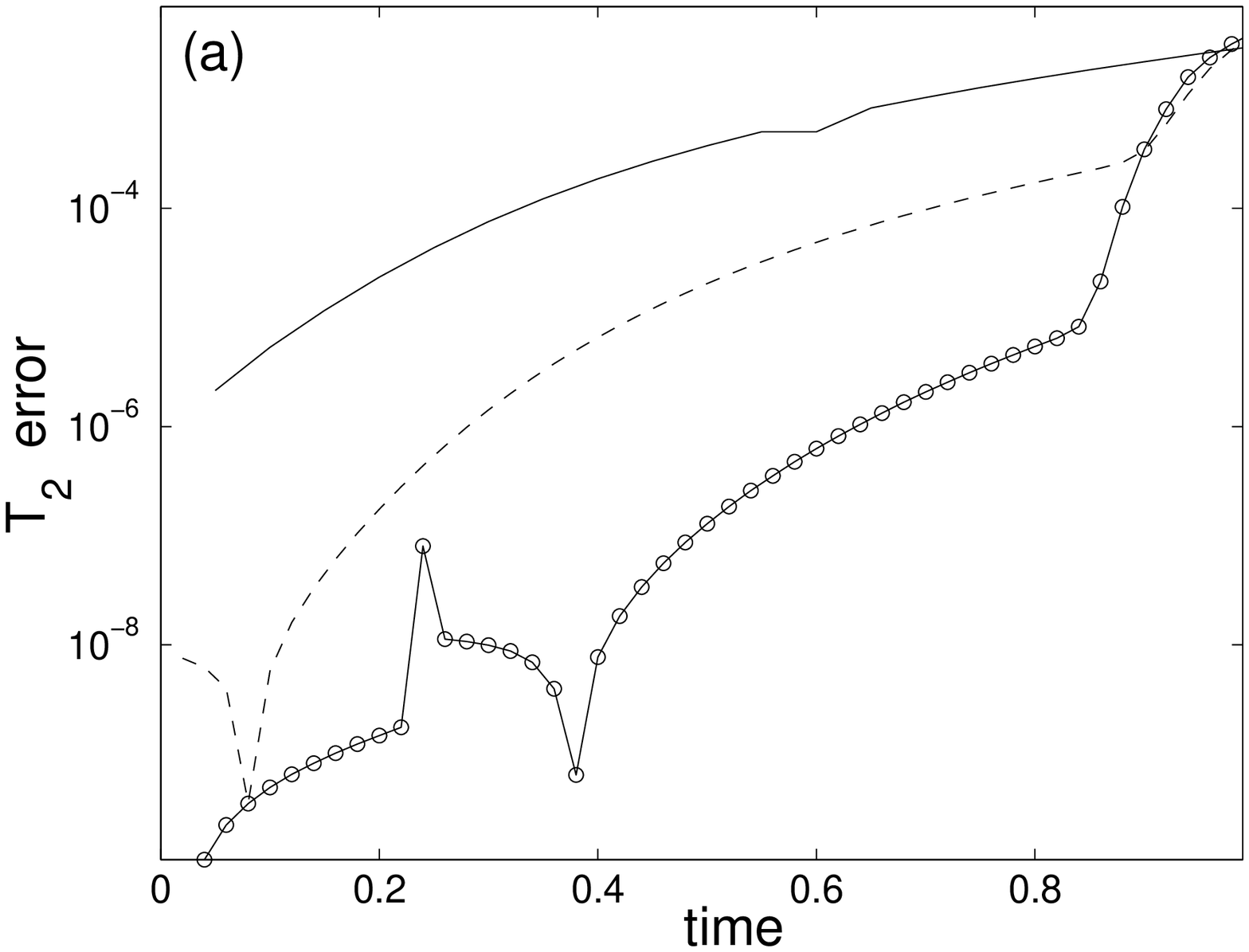}}
\end{minipage}
\begin{minipage}[c]{.3 \linewidth}
\scalebox{1}[1]{\includegraphics[width=\linewidth]{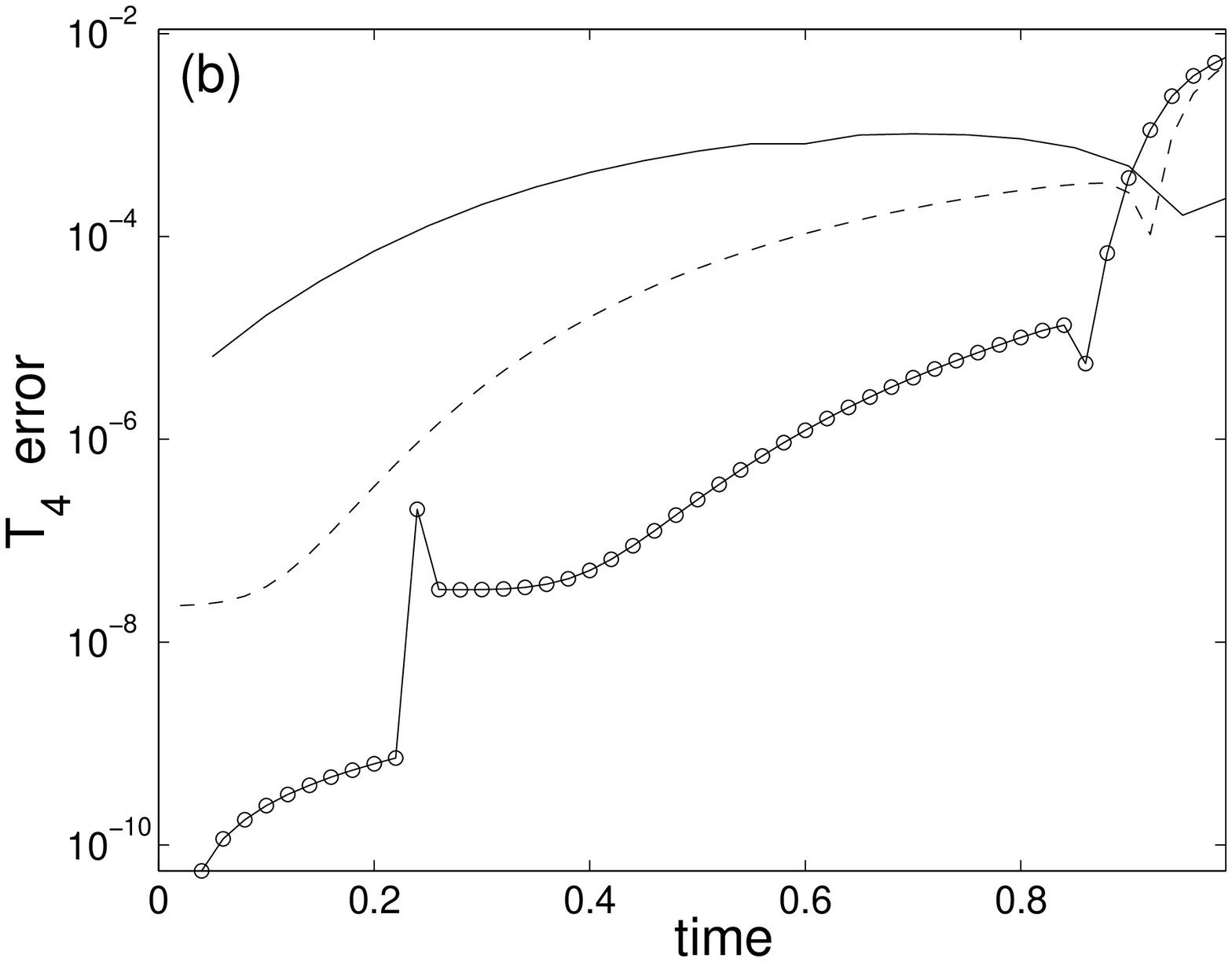}}
\end{minipage}
\begin{minipage}[c]{.3 \linewidth}
\scalebox{1}[1]{\includegraphics[width=\linewidth]{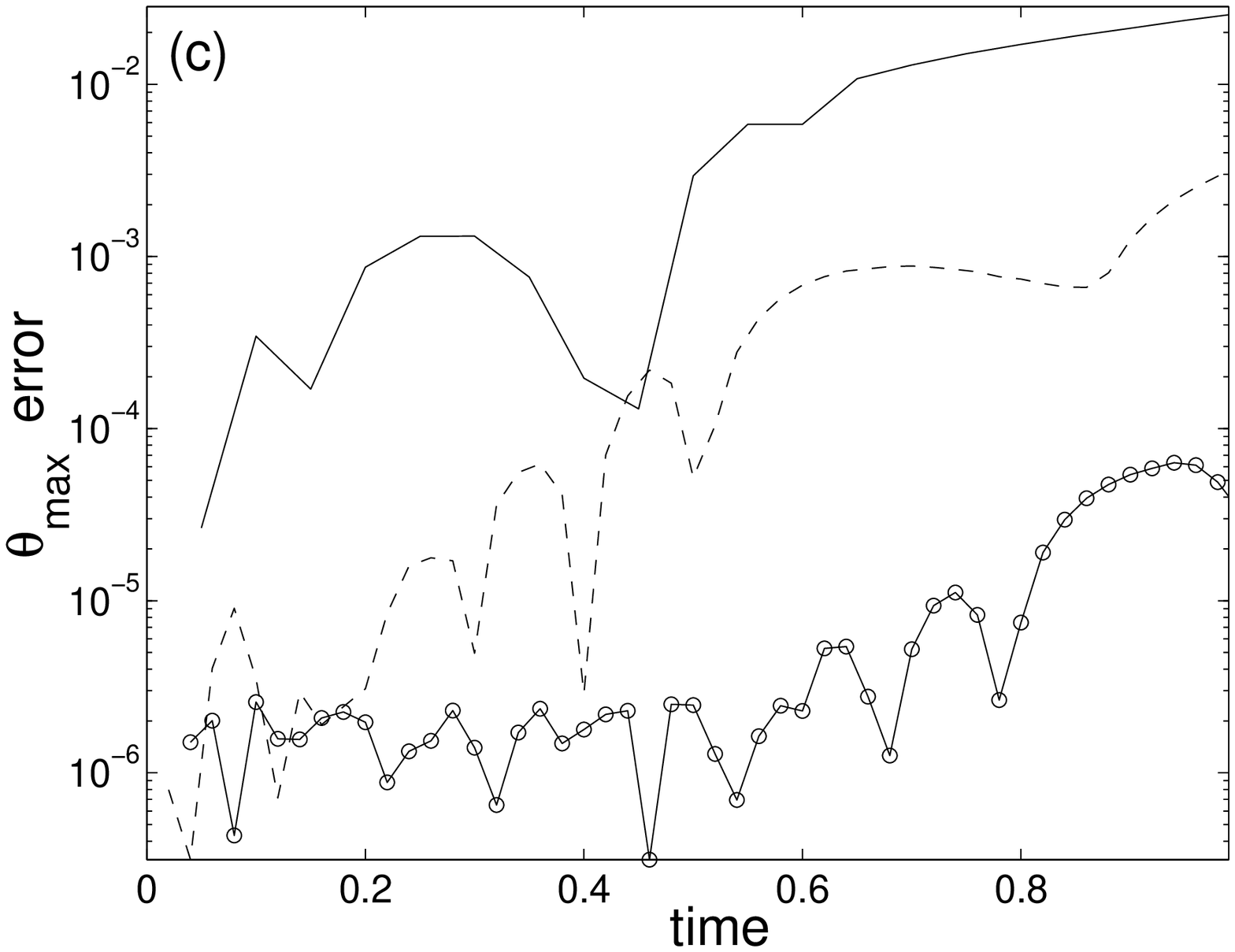}}
\end{minipage}
\caption{\label{fig:error} The evolution of the $T_2$, $T_4$ and
maximum $\theta$ errors for three different resolutions ($1024^2$:
solid line, $2048^2$: dashline, $4096^2$: circle line). The errors
are respectively defined as ${(T_2(0)-T_2(t))}/{T_2(0)}$,
${(T_4(0)-T_4(t))}/{T_4(0)}$ and $|\theta_{max}(0) -
\theta_{max}(t)|/|\theta_{max}(0)|$.}
\end{figure*}

\begin{figure}
\begin{minipage}[c]{.8 \linewidth}
\scalebox{1}[0.9]{\includegraphics[width=\linewidth]{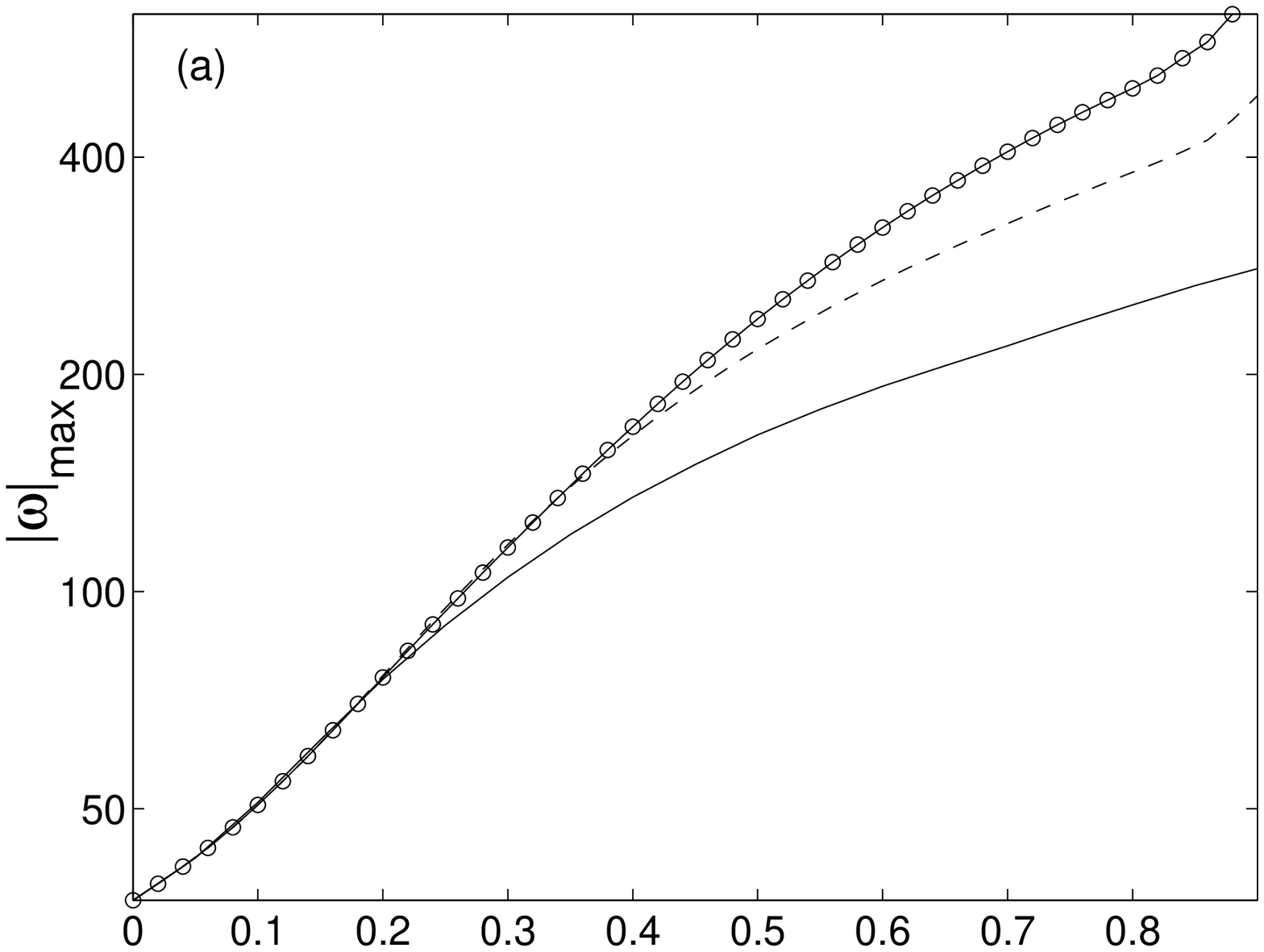}}
\end{minipage}
\begin{minipage}[c]{.8 \linewidth}
\scalebox{1}[0.9]{\includegraphics[width=\linewidth]{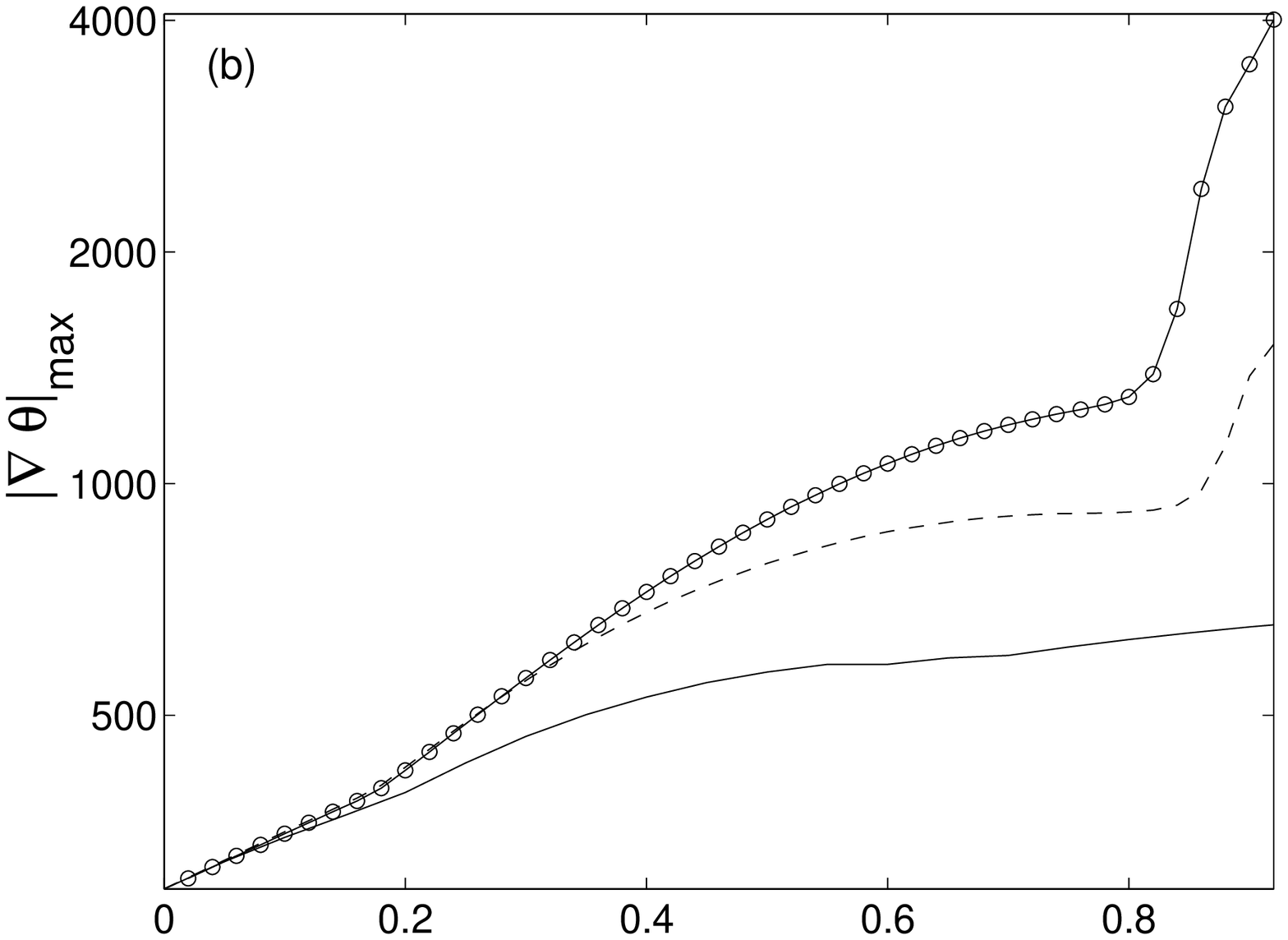}}
\end{minipage}
\caption{\label{fig:max} (a) and (b) are the time evolutions of
maximum $|\omega|$ and $|\nabla \theta|$ ($1024^2$: solid line,
$2048^2$: dashline, $4096^2$: circle line). It seems that $1024^2$
and $2048^2$ runs have almost identical peak values from $t = 0$
until $t = 0.15$, and the results of $2048^2$ and $4096^2$ are
very close until $t = 0.3$. There are two reasons for these
departures: 1) the filter removes more energy for lower
resolution; 2) higher resolution can make a better fit for
delta-like functions. }
\end{figure}

The non-dimensionized 2D inviscid Boussinesq convection equations
can be written in the $\omega$-$\psi $ formulation:
\begin{eqnarray}
&& \theta_t + {\rm {\bf u}} \cdot \nabla \theta = 0, \label{eq23}
\\
&& \omega _t + {\rm {\bf u}} \cdot \nabla \omega = - \theta_x, \label{eq24}
\\
&& \Delta \psi = - \omega, \label{eq25}
\end{eqnarray}
where the gravitational constant
is normalized to $\textbf{g}=(0,-1)$, $\theta$ is the temperature,
${\rm {\bf u}} =
(\mbox{u,v})$ the velocity, $\nu$ the kinematic viscosity, ${\rm {
\mbox{ \boldmath{$\omega$}}}} = (0,0,\omega ) = \nabla \times {\rm
{\bf u}}$ vorticity, and $\psi$ stream function. The simulation is
carried out in the $[0, 2 \pi]^2$ doubly-periodic
domain. At first, we take the initial condition with unified zero
vorticity and a cap-like contour of temperature with the following
expression:
\begin{equation}
\label{eq33} \theta (x,y,0) = (\frac{4x-3\pi}{\pi}) \theta _1
(x,y)\theta _2 (x,y)\left[ {1 - \theta _1 (x,y)} \right],
\end{equation}
where if $S(x, y):= \pi^2 - y^2 - (x - \pi )^2$ is positive,
 then $\theta_1 = \exp {\left( 1 - \pi^2/S(x,y)\right)}$,
and zero otherwise; if $s(y):= \left| {y - 2\pi } \right|
/1.95\pi$ is less than 1, then $\theta_2 = \exp \left( 1- (1-
s(y)^2) ^{-1} \right)$, and zero otherwise. This initial condition
(\ref{eq33}) is similar to \cite{e94} except that a factor
$(4x-3\pi)/{\pi}$ s used to break down the field symmetry with
respect to $x = \pi$. The main flow structure looks like a rising
bubble which touches the $y = 0$ and $y = 2\pi$ boundaries at $t
\simeq 2.0$. This will cause the singularity at the axis if we
transform the 2D Boussinesq results back to the 3D axisymmetric
flow ~\cite{mar02}. To fix this problem, we compress the
intermediate results at $t=1.2$ obtained by using the initial
condition (\ref{eq33}) to form a new initial data. More precisely,
we let $\omega(x,y,0) = \omega'(x,2y-0.4\pi,1.2)$ and
$\theta(x,y,0) = \theta'(x,2y-0.4\pi,1.2)$, for $(x, y)\in [0,
2\pi] \times [0, \pi]$ (where $\theta'$ and $\omega'$ are obtained
by solving (\ref{eq23})-(\ref{eq33}) with a $2048^2$ grid), and
zero otherwise. The new initial condition only needs about 1/4
simulation time to reach $T_c$ comparing with the run in
~\cite{e94}, and advantages of higher resolutions show up at early
stages of simulations (Figs. \ref{fig:max}). The time steps for
the resolutions $1024^2$, $2048^2$ and $4096^2$ are 0.0004, 0.0002
and 0.0001, respectively, given by the CFL condition.

The compressed bubble (Figs. \ref{fig:contour}(a)(d)) will
continue to rise, and  the edge of the cap will roll up at
later times (Figs.
\ref{fig:contour}(c)(f)). At $t \simeq 0.9$, the density and
vorticity contours develop into the shape of ``two asymmetric
eyes.'' The contour revolutions on three different resolutions
reveal similar phenomena at this point. On the other hand, combining the
divergence-free condition, the doubly-periodic condition and  Eq.
(\ref{eq23}), we can verify that $T_2(t) =
\int^{2\pi}_{0}\int^{2\pi}_{0} \theta^2(x,y,t) dxdy$ and $T_4(t) =
\int^{2\pi}_{0}\int^{2\pi}_{0} \theta^4(x,y,t) dxdy$ are
time independent. Figs. \ref{fig:error} (a) and (b) demonstrate that the
global average quantities are well conserved within $1\%$ error
for all the three resolutions used. The flow field develops into many small
vortices after $t = 0.8$ and more and more energy is removed by
the filter in the time evolution. Hence, it seems that
simulations with higher
resolution do not have more advantages here.

However, the time evolution of the maximum values of $\theta$, which is also
time-independent, shows much better
improvement for the $4096^2$ run by comparing with lower
resolution runs (Fig. \ref{fig:error} (c)): the relative error for
$|\theta|_{max}$ with the $4096^2$ run
is always lower than $10^{-4}$, while the corresponding error with the
$1024^2$ run is about $10^{-2}$.
Because the global maximum values
of $|\omega|$ and $|\nabla \theta|$ are normally used as the
indicators in the singularity analysis, the $4096^2$ run may lead
to a more accurate conclusion than that of the $1024^2$ and
$2048^2$ runs. It should be pointed out that if resolutions higher
than $4096^2$ are employed then  the quadruple precision
($\epsilon = 10^{-32}$) instead of the double precision ($\epsilon
= 10^{-16}$) may have to be used; otherwise the results may be
spoiled by the round-off errors (see ~\cite{yin05} for more
details). The maximum vorticity is initially located on the left edge of the
rising bubble (Fig. \ref{fig:contour}(a)), and later moved to the
outer layer of the ``eyes'', (see, Figs. \ref{fig:contour}(c) and
\ref{fig:vorticity}(a). This observation is in good
agreement to the prediction
in ~\cite{e94}. At $t = 0.88$, the $|\omega|_{max}$ locates at
(5.31, 2.72) (see the ``+'' in Fig. \ref{fig:vorticity} (a)).

\begin{figure}
\begin{minipage}[c]{.8\linewidth}
\scalebox{1}[0.9]{\includegraphics[width=\linewidth]{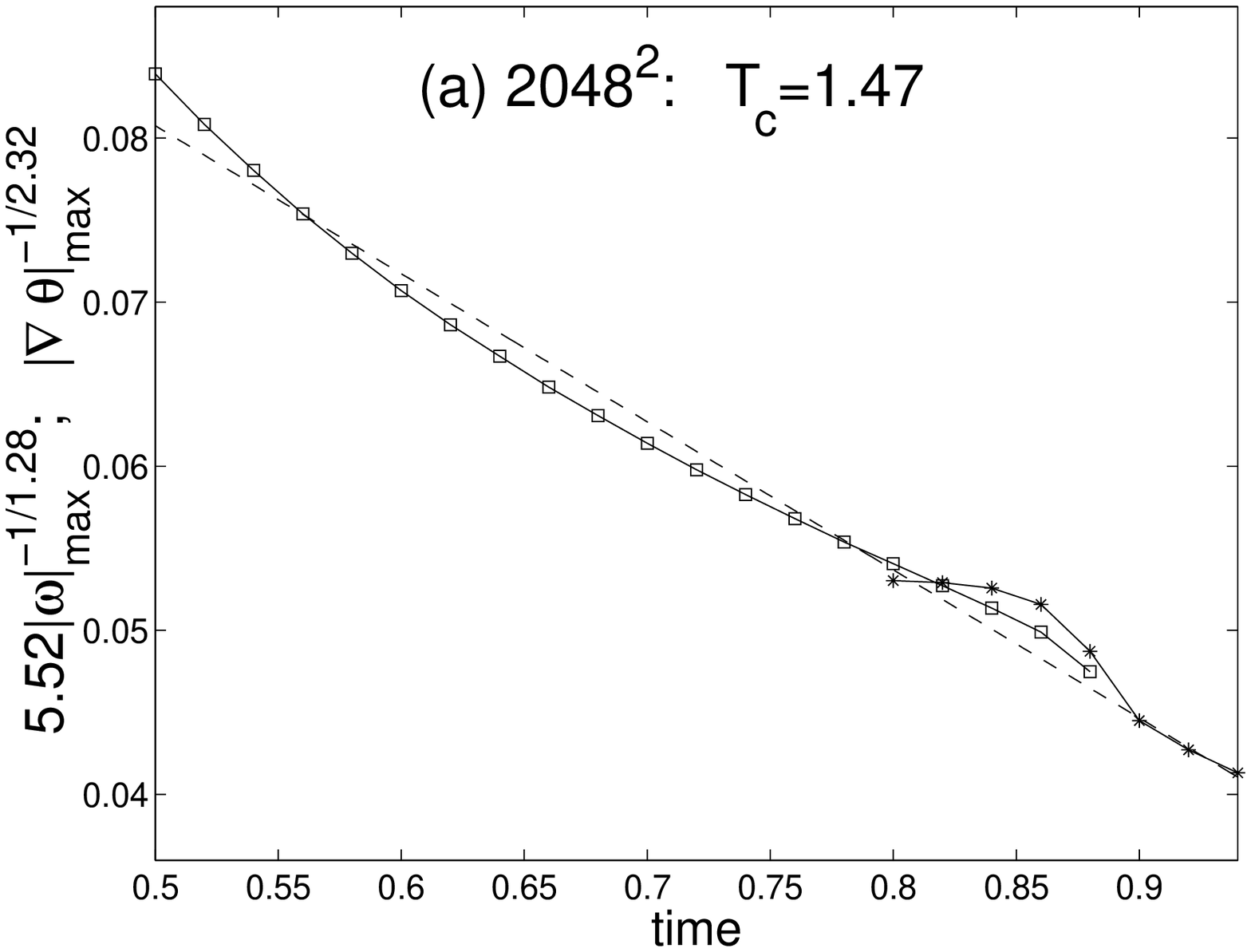}}
\end{minipage}
\begin{minipage}[c]{.8\linewidth}
\scalebox{1}[0.9]{\includegraphics[width=\linewidth]{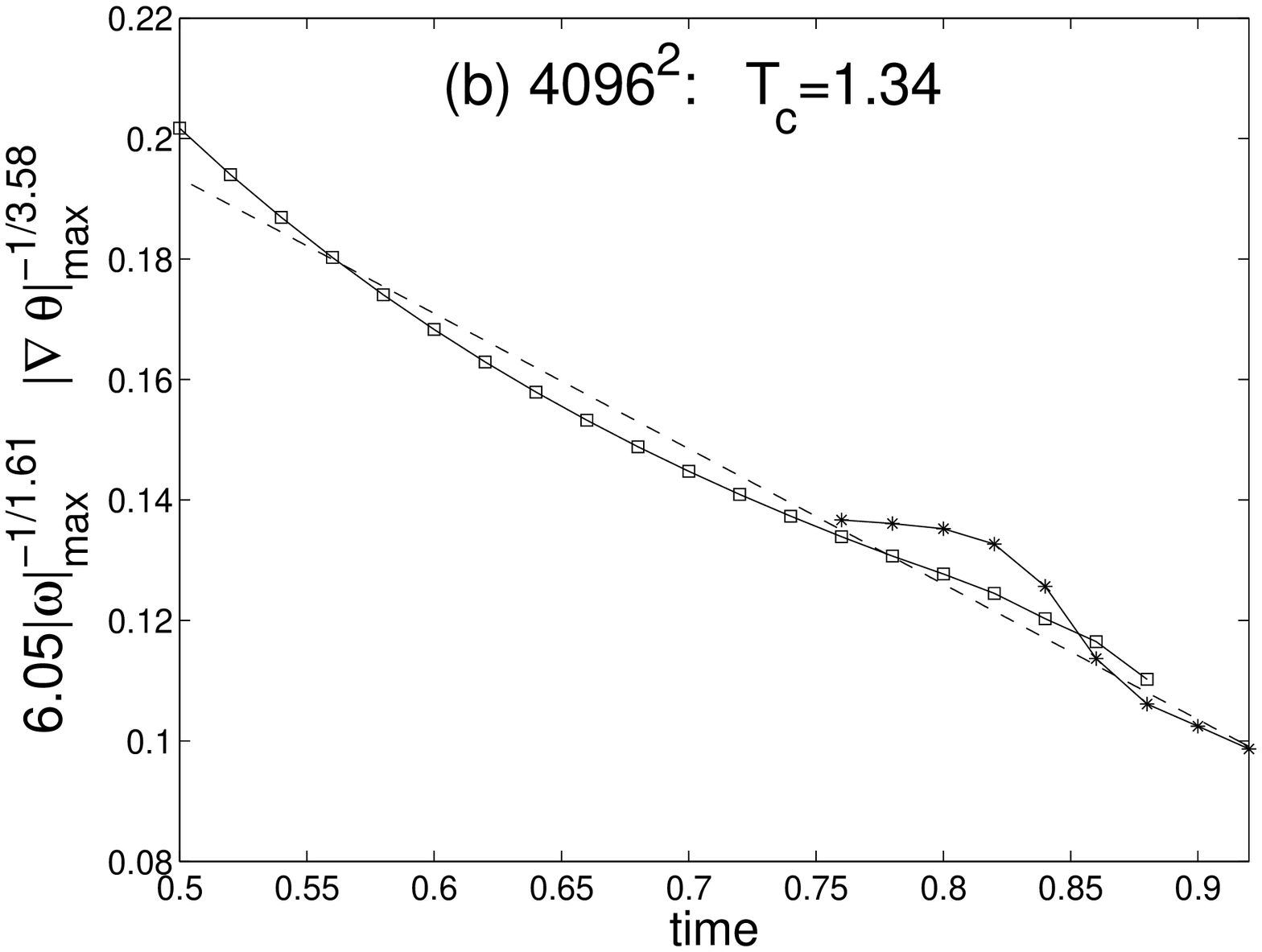}}
\end{minipage}
\caption{\label{fig:rescale} The time evolutions of re-scaled
$|\omega|_{max}$ (square line), $|\nabla \theta|_{max}$ (star
line) and their least square fit to a straight line (dash line)
for the (a) $2048^2$ and (b) $4096^2$ run. The blow-up times
($T_c$) are the points where x-axes are intersected by dash lines.
The $1024^2$ run has not been plotted because there is no blow-up
signal until $t = 1.0$.}
\end{figure}

Around $t \simeq 0.88$, some very small vorticity structures begin
to appear at the lower part of the smooth outer layer where the
maximum $|\omega|$ appears. The filter we adopted in the code will
remove more and more energy from the system, so that the global
average values like $T_2$ and $T_4$ are greatly affected (Figs.
\ref{fig:error} (a)(b)). Consequently, after this critical time
the numerical results in the $4096^2$ run may not be accurate and
reliable.  Actually, it is observed that the $|\omega|_{max}$ and
$|\nabla \theta|_{max}$ experience a drop-down after $t \simeq
0.9$. This indicates that the filter prevents the blowup of the
maximum vorticity from happening, which is similar to the
viscosity effect in high Reynolds number simulations (e.g.
~\cite{bor94,ker05}). Therefore, the sample maximum values after
the dropping down should not be used in the singularity analysis.
For the $1024^2$ and $2048^2$ runs, however, the drop-downs appear
later than $t \simeq 0.88$ and the blow-up time $T_c$ is also
delayed. In Figs. \ref{fig:max}, we only provide time evolutions
for $|\omega|_{max}$ and $|\nabla \theta|_{max}$ in different
resolutions before they drop down. We re-plot these maximum value
evolutions in Fig. \ref{fig:rescale}. It seems that the growth of
$|\omega|_{max}$ and $|\nabla \theta|_{max}$ in the $4096^2$ run
(Fig. \ref{fig:rescale}(b)) plausibly terminates in a finite time
with $|\omega|_{max} \sim {(T_c-t)}^{-1.61}$, and crudely $|\nabla
\theta|_{max} \sim {(T_c - t)}^{-3.58}$ (in the later case we only
adopt the values of $|\nabla \theta|_{max}$ after $t = 0.75$). It
is noticed that our values  $(\alpha, \beta) =(1.61, 3.58)$ are
larger than the minimum blow-up values ~\cite{e94, cha97} and the
existing results~\cite{pum92a,pum92b,e94}. We run our simulations
until $t = 2.0$ and find that the main structure of the flow field
stays away from $y = 0$ and $y = 2\pi$. The axisymmetric
assumption adopted in the original 3D Euler equation model does
not cause any axis singularity difficulty as discussed earlier
with our new  initial condition.

We have not performed the simulation on grid finer than $4096^2$,
but we can predict some results from the present computations. It
is well known that when a delta function is approximated by a
finite Fourier series, each doubling of the resolutions will cause
doubling of the maximum value and $2^{n+1}$ times the maximum
values of the $n^{th}$ order space derivative (see the analysis in
Appendix A of ~\cite{bor94}). In the final state of our simulation
(Fig. \ref{fig:vorticity}(a)), the cut-line at $y = 2.72$ through
the out layer of the ``eye''  looks very similar to a delta
function. Therefore, when finer and finer resolutions are used the
peak vorticity and temperature gradient are getting larger and
larger. Further 2D Boussinesq simulations with even higher
resolution will support our singularity prediction although it
seems impossible to accurately predict the singularity time with
the current computational scheme. In Fig. \ref{fig:max}(b), the
gradients of the curves become larger when the grids are refined.
The $1024^2$ result indicates no blow-up, and the singularity
analysis for the $2048^2$ run (Fig. \ref{fig:rescale}(a)) shows
that $\alpha = 1.28$, $\beta = 2.32$ and $T_c = 1.47$. The results
obtained by the three resolutions reveal that the values of $T_c$
may be smaller than 1.34, and the values of $\alpha$ and $\beta$
may be larger than 1.61 and 3.58 if much fine resolutions ($>
4096^2$) are used. Although our simulations can not provide
accurate value for $T_c$, the existence of singularity in the 2D
Boussinesq simulations is strongly supported by our numerical
computations.

We would like to thank Prof. Linbo Zhang for the support of using
the local parallel computers. This work is supported by National
Natural Science Foundation of China (G10476032). TT thanks the
supports from International Research Team on Complex System of
 Chinese Academy of Sciences and from Hong Kong
Research Grant Council.

\end{document}